\documentclass[useAMS,usenatbib]{mn2e}
\usepackage{graphicx}

\title[Blow-Off of ``Super-Earth'' Atmospheres]
  {Probing the Blow-Off Criteria of Hydrogen-Rich ``Super-Earths''}
\author[H. Lammer et al.]
  {H.~Lammer,$^1$ N. V.~Erkaev,$^2$  P. Odert,$^{1,3}$ K. G.~Kislyakova,$^{1,3}$
\newauthor    M. Leitzinger$^3$, M. L. Khodachenko,$^1$
 \\
  $^1$Space Research Institute, Austrian Academy of Sciences, Schmiedlstr. 6, A-8042, Graz, Austria\\
  $^2$Institute for Computational Modelling, Russian Academy of Sciences, and Siberian Federal University,\\
    Krasnoyarsk 36, Russian Federation\\
  $^3$Institute of Physics, University of Graz, Universit\"{a}tsplatz 5, A-8010 Graz, Austria}
\date{Released 2012 Xxxxx XX}

\pagerange{\pageref{firstpage}--\pageref{lastpage}} \pubyear{2012}

\def\LaTeX{L\kern-.36em\raise.3ex\hbox{a}\kern-.15em
    T\kern-.1667em\lower.7ex\hbox{E}\kern-.125emX}

\begin{document}

\label{firstpage}

\maketitle

\begin{abstract}
The discovery of transiting ``super-Earths'' with inflated radii and known masses such as Kepler-11b-f,
GJ 1214b and 55 Cnc e, indicates that these exoplanets did not lose their
nebula-captured hydrogen-rich, degassed or impact-delivered protoatmospheres by atmospheric escape
processes. Because hydrodynamic blow-off of atmospheric hydrogen atoms is the most efficient
atmospheric escape process we apply a time-dependent numerical algorithm which is able to
solve the system of 1-D fluid equations for mass, momentum, and energy conservation to
investigate the criteria under which ``super-Earths'' with hydrogen-dominated upper atmospheres
can experience hydrodynamic expansion by heating of the stellar XUV (soft X-rays and extreme ultraviolet)
radiation and thermal escape via blow-off. Depending on orbit location, XUV flux, heating efficiency
and the planet's mean density our results indicate that
the upper atmospheres of all ``super-Earths'' can expand to large distances,
so that besides of Kepler-11c all of them experience atmospheric mass-loss due to Roche lobe
overflow. The atmospheric mass-loss of the studied ``super-Earths'' is
one to two orders of magnitude lower compared to that of ``hot Jupiters'' such as
HD 209458b, so that one can expect that these exoplanets cannot lose their
hydrogen-envelopes during their remaining lifetimes.
\end{abstract}


\section{INTRODUCTION}
The detection of hydrogen- and volatile-rich exoplanets at orbital distances $<$1 AU
opens questions regarding their upper atmosphere structures and the stability against
escape of atmospheric gases. Since $\geq$40\% of all discovered exoplanets are orbiting their host
stars at distances
closer than the orbit of Mercury, the atmospheres of these bodies evolve in much more extreme environments
than what is known from the planets in our Solar System. More intense stellar X-ray, soft X-ray, EUV\footnote{The radiation between
5-920 $\rm \AA$ contains soft X-rays and EUV and is considered as XUV radiation} radiation
and particle fluxes at such close orbital distances will alter the upper atmosphere structure of these
objects to a great extent.

Lammer et al. (2003) were the first to provide a model of
exoplanetary upper atmospheres, which is based on an approximate solution of the heat
balance equation in planetary thermospheres (Bauer 1971; Gross 1972) and found
that hydrogen-rich upper atmospheres of Jupiter-type gas giants in close orbital
distances will be heated to several thousands of Kelvin, so that hydrostatic conditions
cannot be valid anymore, because the planet's upper atmosphere will expand dynamically
upwards. Under such conditions the exobase can move to locations which are above a possible
magnetopause or even at the Roche lobe distance (e.g., Lammer et al. 2003; Lecavelier des Etangs et al. 2004;
Erkaev et al. 2007; Lammer et al. 2009a), so that the upward-flowing neutral gas can
interact with the dense stellar plasma flow (Holmstr\"{o}m et al. 2008; Ekenb\"{a}ck et al. 2010;
Lammer et al. 2011). Hydrodynamic models by Yelle (2004), Tian et al. (2005a), Garc\'{\i}a Mu\^{n}oz (2007),
Penz et al. (2008), and Koskinen et al. (2010) agreed also with the hypothesis of Lammer et al. (2003; 2009a)
that close-in hydrogen-rich exoplanets experience dynamic atmospheric expansion and outflow up to their Roche lobes
with mass-loss rates in the order of $\sim(1-5)\times10^{10}$ g s$^{-1}$.

The first observational evidence of an XUV-heated extended non-hydrostatic
upper atmosphere was obtained for the hot gas giant HD 209458b, which orbits a Sun-like star at
0.047 AU (Vidal-Madjar et al. 2003; Ben-Jaffel 2007; Ben-Jaffel \& Sona Hosseini 2010).
Excess absorption in the Lyman-$\alpha$ line of $\sim$9--15 \% during the transit was interpreted as
being caused by a population of atomic hydrogen located beyond the Roche lobe, indicating a hot and extended
upper atmosphere. Subsequent detection of O I and C II by Vidal-Madjar et al. (2004), as well as Si III by
Linsky et al. (2010), supports the hypothesis that these heavy elements are likely to be carried
there from deeper atmospheric levels with the dynamical hydrogen flow. Metals have been found
around the Roche lobe of the highly irradiated ``hot Jupiter'' WASP-12b, an inflated gas giant at $\sim$0.023 AU,
by Fossati et al. (2010). As heavier elements can only be dragged to the upper atmosphere regions
along with upward flowing hydrogen atoms, these observations of heavy elements at large distances from the
planet most likely confirm scenarios which were suggested by Sekiya, Nakazawa, \& Hayashi (1980),
Sekiya,  Hayashi, \& Nakazawa (1981), Hunten, Pepin, \& Walker (1987) and
Pepin (1991; 2000) that hydrodynamic escape from early Solar System planets like Venus could
have generated Ne and Ar isotope ratios close to the observed values in its present-time atmosphere and
noble gas ratios similar to those derived for Earth's initial atmosphere. In view of atmospheric isotope
studies it is generally accepted that hydrodynamic outflow driven by the XUV-active young Sun
could have been responsible for the observed heavy isotope enrichment in the atmospheres of Venus,
Earth and Mars (e.g., Lammer et al. 2008; and references therein).

Therefore, depending on the activity of a planet's host star, its orbit location, mass and size,
one can expect that some exoplanets experience high atmospheric mass-loss during their lifetimes
(Lammer et al. 2003; Vidal-Madjar et al. 2003; 2004; Lecavelier des Etangs et al. 2004; 2010;
Lecavelier des Etangs 2007; Yelle 2004; Baraffe et al. 2004; Tian et al. 2005a; Erkaev et al. 2007;
Hubbard et al. 2007; Khodachenko et al. 2007; Garc\'{\i}a Mu\^{n}oz 2007; Koskinen, Aylward, \& Miller 2007;
Penz et al. 2008; Penz, Micela, \& Lammer 2008; Penz \& Micela 2008; Davis \& Wheatley 2009; Lammer et al. 2009a;
Murray-Clay, Chiang, \& Murray 2009; Linsky et al. 2010; Fossati et al. 2010; Leitzinger et al. 2011; Owen \& Jackson 2012).

In this context, the discoveries of very close-in low-mass exoplanets like CoRoT-7b (e.g. L\'{e}ger et al. 2009),
and Kepler-10b (e.g., Batalha et al. 2011) at $< 0.02$ AU raise the question whether
these objects have always been so-called ``super-Earths'', or if are they just remnants of
once more massive gaseous planets which lost their whole hydrogen envelopes (e.g., Valencia et al. 2010;
Jackson et al. 2010; Leitzinger et al. 2011). However, besides rocky ``super-Earths'' which orbit at
very close orbital distances such as CoRoT-7b, Kepler-10b, Kepler-18b (Cochran et al. 2011)
and planets in the Kepler-20 system (Fressin et al. 2012),
the Kepler space observatory discovered several low-density ``super-Earths'' closely packed in the Kepler-11
system (Lissauer et al. 2011), whose mean densities indicate that they have rocky cores which are surrounded
by significant amounts of hydrogen/helium envelopes (e.g., Lissauer et al. 2011; Lammer et al. 2012; Lammer 2012;
Ikoma \& Hori 2012). Moreover, the mean densities of other transiting ``super-Earths'', such as GJ 1214b (Charbonneau et al. 2009)
or 55 Cnc e (Demory et al. 2012; Endl et al. 2012; Gillon et al. 2012), also indicate substantial envelopes of light gases such as hydrogen
and He or H$_2$O. For GJ 1214b, the current status of observational evidence supports mainly two possible atmospheres,
namely a H/He-dominated atmosphere with clouds and low methane content, or a H$_2$O-dominated atmosphere
(Bean et al. 2010, 2011; Croll et al. 2011; D\'esert et al. 2011; de Mooij et al. 2012). For the large ``super-Earth''
55 Cnc e three different composition-based hypotheses can be found in the literature, within the first two possibilities are
a rocky planet with a H/He dominated atmosphere, or a supercritical water planet (Demory et al. 2012; Gillon et al. 2012).
The third hypothesis is based on a recent study by
Madhusudhan et al. (2012) where these authors suggest that the mass and radius of 55 Cnc e can also be explained by a carbon-rich solid
interior made of Fe, C, SiC, and/or silicates and without a volatile envelope. Although, there may be this possibility of
an alternative more exotic explanation for 55 Cnc e, the discoveries of these low-density hydrogen and/or H$_2$O-rich
planets can be seen as an indication that many ``super-Earths'' even located close to their host stars will not lose their
initial volatile-rich protoatmospheres during their lifetimes (Ikoma \& Hori 2012; Lammer 2012). If so, the consequences of these findings
are very relevant for the probability and evolution of Earth-type class I habitats (Lammer et al. 2009b; Lammer 2012)
and habitability in general (Pierrehumbert \& Gaidos 2011; Wordsworth 2012). Understanding the efficiency of atmospheric loss processes at
hydrogen-rich ``super-Earths'' is therefore crucial.

The aim of this study is to investigate the environmental conditions under which
``super-Earths'' with hydrogen-rich upper atmospheres undergo hydrodynamic blow-off. For this study we focus only on ``super-Earths''
for which the sizes and masses are more or less well determined. For that reason we do not include transiting ``super-Earths''
such as HD 97658b, Kepler-9d, or small terrestrial planets such as those discovered in the
Kepler-42 system with unknown or uncertain masses, or ``super-Earths'' with known masses but unknown radii.

In Sect. 2 we describe the adopted stellar and planetary input parameters and the hydrodynamic upper atmosphere model
which is applied for the investigation if Kepler-11b, Kepler-11c,
Kepler-11d, Kepler-11e, Kepler-11f, GJ 1214b and experience atmospheric blow-off or less efficient Jeans escape.
In the controversial case regarding the composition of 55 Cnc e, we assume in this study
that the planet is surrounded by a hydrogen envelope, which would be the case if rocky core of the planet is surrounded by
a H/He, or H$_2$O dominated atmosphere as suggested by Demory et al. (2012) or Gillon et al. (2012).

In Sect. 3 we apply our hydrodynamic model to these low-density ``super-Earths''
and study the response of their upper atmospheres to the stellar XUV flux and calculate the thermal atmospheric
escape rates. We compare our results to the widely used energy-limited blow-off formula and discuss
the relevance of our findings for ``super-Earths'' in general.
\section{INPUT PARAMETER AND MODEL DESCRIPTION}
\subsection{XUV fluxes of Kepler-11, GJ 1214 and 55 Cnc}
Unfortunately, the XUV fluxes of the host stars of the studied ``super-Earths''
are not very well constrained. The XUV emission of Kepler-11, a slightly evolved late-G star (Lissauer et al. 2011),
is observationally unconstrained, because its large distance ($>600$\,pc) and high age (6--10\,Gyr) prevent
detection by current instruments. Hence, we estimate its XUV flux via its age by using power laws from Ribas et al. (2005),
although we caution that application of these calibrations for stellar ages higher than about 6.7\,Gyr (the highest age in their sample)
might introduce additional uncertainties because of the extrapolation to an even older, slightly evolved G star.
Adopting a mean age of 8\,Gyr, this approach yields a $\log L_\mathrm{XUV}\approx27.81\,\mathrm{erg\,s^{-1}}$, which corresponds to fluxes of
$F_\mathrm{XUV}\sim 278, 204, 91, 61, 37\,\mathrm{erg\,cm^{-2}\,s^{-1}}$ at the respective orbits of the Kepler-11 planets b-f.
We estimate the uncertainties of these values to be at least an order of magnitude due to the extrapolation of the calibrations
and the just loosely constrained stellar age.

Although GJ~1214 is located at a closer distance ($\sim$13\,pc), the star has not been detected in X-rays up to now.
This is probably due to the intrinsic faintness of this M4.5V star and its age of 3--10\,Gyr (Charbonneau et al. 2009).
Therefore, we use the relation between X-ray luminosity and age of Engle \& Guinan (2011) calibrated for M dwarfs.
This relation yields an X-ray luminosity of $\log L_\mathrm{X}\approx26.8\,\mathrm{erg\,s^{-1}}$ for an adopted age of 6\,Gyr.
To correct for the unknown EUV emission, we assume that the contributions to the total XUV flux of both X-rays and EUV are $\sim$50\% each,
typical for moderately active stars (e.g. Sanz-Forcada et al. 2011), but EUV can make up to 90\% of the emission for inactive stars.
This leads to an estimated XUV flux at GJ~1214b's orbit of $F_\mathrm{XUV}\approx2200\,\mathrm{erg\,cm^{-2}\,s^{-1}}$.
As for Kepler-11, the uncertainty of this value also corresponds at least an order of magnitude because of the large range of
possible stellar ages, the intrinsic spread of about one order of magnitude of coeval stars of the same mass, and the very
crude correction applied for the unknown EUV emission.

For 55 Cnc, a nearby ($\sim$12.5\,pc) G9IV star (von Braun et al. 2011),
X-ray emission was detected by ESA's XMM-Newton space observatory (Sanz-Forcada et al. 2011). These authors
used X-ray spectra of numerous planetary host stars to extrapolate their EUV emission by using calibrations
of stars with well-determined emission measure distributions (EMD). They estimated a total XUV luminosity
within the wavelength region 5--920\AA\ of $\log L_\mathrm{XUV}=27.55^{+0.46}_{-0.42}\,\mathrm{erg\,s^{-1}}$,
which translates to an XUV flux of $F_\mathrm{XUV}\approx4913\,\mathrm{erg\,cm^{-2}\,s^{-1}}$ at the orbit
of 55~Cnc~e. The uncertainty is $\geq 50\%$, mainly due to the uncertainties in the extrapolated transition region
EMD. Although X-rays might contribute somewhat to the heating of the lower thermosphere,
from our analysis of the available data we assume that for the particular atmospheric escape of the investigated
planets is mainly driven by the stellar EUV emission. This assumption is supported by the fact that
for rather old stars like the ones studied here, however, the major portion of the XUV flux (80--90\%)
is emitted in the EUV range.

The properties of the studied planets are summarized in Table 1,
which also gives the XUV enhancement factors $I_{\rm XUV}$, corresponding to the ratio of the adopted
XUV fluxes at the planet's orbits
normalized to the present XUV flux of the Sun at 1\,AU ($4.64\,\mathrm{erg\,cm^{-2}\,s^{-1}}$; Ribas et al. 2005).
\subsection{Energy absorption and model description}
The energy budget of upper planetary atmospheres is mainly governed by the heating of
the bulk atmosphere due to the absorption of the XUV radiation (e.g. Bauer \& Lammer 2004) and in
very close orbital distances also harder X-rays (Cecchi-Pestellini et al. 2009;
Owen \& Jackson 2012) by atmospheric constituents, by heat transport due to conduction and convection and by heat
loss due to emissions in the infrared (IR). Because the XUV fluxes are based only rough estimates except for
55 Cnc e, and the atmospheric composition of the studied planets is not very well constraint, we do not consider
a wavelength dependent derivation of the heating function. But we note that
for future studies of exoplanets with known XUV spectra and atmospheric composition
it might be important to investigate the influence of wavelength dependent absorption cross sections
and their impact on the upper atmosphere structure.
The volume heat production rate $q_{\rm XUV}$ due to the absorption of
the stellar radiation for given wavelengths and
constituents can then be written as
\begin{equation}
q_{\rm XUV} = \eta n \sigma_{\rm a}J_{\rm XUV}e^{-\tau},
\end{equation}
where $J_{\rm XUV}$ is the energy flux related to the corresponding wavelength range outside
the atmosphere, $\sigma_{\rm a}$ is the appropriate hydrogen absorption cross-section which lies typically
within the range of $\sim$ $10^{-18}$--$6\times 10^{-18}$\,cm$^2$) and $\tau$ is the
optical depth in the upper atmosphere and $n$ the number density.
$\eta$ is the so-called heating efficiency or fraction of absorbed stellar
radiation which is transformed into thermal energy and lies within
in the range of $\sim$15--60 \% (Chassefi\`{e}re 1996a; 1996b; Yelle 2004; Lammer et al. 2009; Leitzinger et al. 2011;
Koskinen et al. 2012).

The value of $\eta$ is also related to the availability of IR-cooling molecules such
as CO$_2$, H$_3^+$, etc. and will then be closer to the lower value because less energy can be
transferred into heat due to cooling by IR-emitting molecules.
In a recent study by Koskinen et al. (2012), who applied a thermosphere model that
calculates the heating rate based on the absorption of stellar XUV
radiation and photoelectron heating efficiencies and photochemistry to the hydrogen-rich
gas giant HD 209458b, it was found that $\eta$ is most likely in the
order of $\sim$40--60 \%. For hydrogen-rich exoplanets which are exposed to XUV fluxes
$\geq$100 times higher compared to that of the present Sun most molecules in the
thermosphere are dissociated so that IR-cooling becomes less important yielding $\eta\sim$40--60 \%.
As this is the case for HD 208459b, IR-cooling by H$_3^+$ molecules can be neglected (Koskinen et al. 2007).
However, for the Kepler-11 ``super-Earths'' which are exposed to XUV fluxes $\sim$8--60 times
higher than today's Sun, depending on the availability of potential IR-cooling
molecules, $\eta$ may be closer to the lower value of 15\%.

As mentioned in Sect. 2, it is also necessary to check if escape by X-ray heating
could be relevant for the studied ``super-Earths''.
However, as X-ray heating usually dominates at young stellar ages (Owen \& Jackson 2012),
it is unlikely to be of relevance
here because of the rather high ages of the host stars. If we compare the X-ray luminosities of the host stars
with the values necessary for having dominating X-ray escape (Owen \& Jackson 2012; their Fig. 11), one can estimate
that X-ray luminosities of at least $10^{28}$ or $10^{30}\,\mathrm{erg\,s^{-1}}$ would be necessary for the
close-in GJ 1214b and 55 Cnc e ``super-Earths'' and the further separated Kepler-11b-f planets, respectively,
if we adopt the values for a Neptune-mass planet with a density of $1\,\mathrm{g\,cm^{-3}}$ as a lower limit.
The X-ray luminosities of all studied  ``super-Earth'' host stars are all well below these values
($<10^{27}\,\mathrm{erg\,s^{-1}}$). Therefore, we assume that X-ray induced evaporation can
be neglected or may play a minor role for the studied ``super-Earths'', and XUV radiation should be
responsible for the main heating processes in their hydrogen-rich thermospheres.

By averaging the XUV volume heating rate over the planet's dayside one obtains
\begin{equation}
q_{\rm XUV}(r)=\frac{\eta n \sigma_{\rm a}}{2}\int_{0}^{\frac{\pi}{2}+arccos(\frac{1}{r})} J_{\rm XUV}(r,\Theta)
sin(\Theta)d\Theta,
\end{equation}
with the polar angle $\Theta$ and $J_{\rm XUV}(r,\Theta)=J_{\rm XUV}e^{-\tau}$.
Because the thermal escape rate is determined by the total energy deposited in the upper atmosphere,
which is given by the integral of the XUV volume heating rate  over the radial distance
within the simulation domain, the shape of $q_{\rm XUV}$ has a minor effect to the atmospheric loss rate.
Besides the uncertainties in the
availability of possible thermospheric IR-coolers, other parametrical uncertainties
such as the chosen XUV flux, a possible X-ray heating contribution, the stellar age, etc. are
most likely within the investigated limits of the assumed 15--40 \% $\eta$ range.

For studying the response of the hydrogen-dominated upper
atmospheres of our ``super-Earths'' to the stellar XUV flux
we apply a non-stationary 1D hydrodynamic upper atmosphere model which is a further developed
model based on Penz et al. (2008), and solves the system of the 1-D fluid
equations for mass, momentum, and energy conservation in spherical coordinates.
\begin{equation}
\frac{\partial \rho r^2}{\partial t} + \frac{\partial \rho v
r^2}{\partial r}= 0,
\end{equation}
\begin{eqnarray}
\frac{\partial \rho v r^2}{\partial t} + \frac{\partial \left[ r^2 (\rho v^2+P)\right]}{\partial r} =\rho g r^2 + 2P r,\\
\frac{\partial r^2\left[\frac{\rho v^2}{2}+\frac{P}{(\gamma-1)}\right]}{\partial t}
+\frac{\partial v r^2\left[\frac{\rho v^2}{2}+\frac{\gamma P}{(\gamma - 1)}\right]}{\partial r}=\nonumber\\
\rho v r^2 g + q_{\rm XUV} r^2,
\end{eqnarray}
with pressure
\begin{equation}
P=\frac{\rho}{m_{\rm H}} k T,
\end{equation}
and gravitational acceleration,
\begin{equation}
g=-\nabla\Phi,
\end{equation}
\begin{eqnarray}
\Phi=-G\frac{M_{\rm pl}}{r}-G\frac{M_{\rm star}}{(d-r)}-G\frac{M_{\rm pl}+M_{\rm star}}{2 d^3}\nonumber\\
\left (\frac{M_{\rm star}d}{M_{\rm star}+M_{\rm pl}}-r\right)^2.
\end{eqnarray}
Here, $r$ corresponds to the radial distance from the planetary center, $\rho$, $v$, $P$ and $T$ are
the mass density, radial velocity, pressure and temperature of the bulk atmosphere, respectively. $m_{\rm H}$ is the
mass of the hydrogen atoms, $k$ is the Boltzmann constant, $G$ is Newton's gravitational constant,
and $\gamma$ is the polytropic index or the ratio of the specific heats.

For exoplanets at very close orbital distances one cannot neglect gravitational effects
which are related to the Roche lobe (Erkaev et al. 2007; Penz et al. 2008; Lammer et al. 2009a).
Although, most of our studied ``super-Earths'' have orbital locations closer than 0.1 AU, due to their
low gravity compared to Jovian-type planets their upper atmospheres may expand to several
planetary radii and probably reach the Roche lobe. Therefore, we include in our study also the Roche lobe
related gravitational force, which we refer to as Roche lobe effects with planetary mass $M_{\rm pl}$ is the
and mass of the host star $M_{\rm star}$ and orbital distance of the planet $d$.
\begin{table*}
\renewcommand{\baselinestretch}{1}
\caption{Planetary and stellar model input parameters for Kepler-11b-f (Lissauer et al. 2011;
http://kepler.nasa.gov/Mission/discoveries/), GJ 1214b (Charbonneau et al. 2009) and 55 Cnc e (Demory et al. 2012; Endl et al. 2012; Gillon et al. 2012).
$I_{\rm XUV}$ corresponds to the stellar XUV flux at the planetary orbits normalized to the
solar value at 1 AU.}
\begin{center}
\begin{tabular}{l|cccccc}
Exoplanet & $d$ [AU] & $I_{\rm XUV}$ & $R_{\rm pl}$ [$R_{\rm \oplus}$] & $M_{\rm pl}$ [$M_{\rm \oplus}$] & $\rho_{\rm pl}$ [g cm$^{-3}$] & $T_{0}$ [K]\\\hline
Kepler-11b &   0.091  & 60   & 1.97   & 4.3     & 3.1 &     $\sim$900 \\
Kepler-11c &   0.1   & 45   & 3.15   & 13.5    &  2.3 &    $\sim$833  \\
Kepler-11d &   0.159  & 20   & 3.43   & 6.1     & 0.9 &     $\sim$692 \\
Kepler-11e & 0.195    & 13   & 4.52   & 8.4     & 0.5 &     $\sim$617 \\
Kepler-11f & 0.25     & 8   & 2.61   & 2.3     &  0.7 &    $\sim$544  \\
GJ 1214b   & 0.014    & 470   & 2.67   & 6.55    & 1.87 &     $\sim$475  \\
55 Cnc e   & 0.0156   & 1060   & 2.17    & 8.37    & 4.56  &   $\sim$2360 \\
\end{tabular}
\end{center}
\normalsize
\end{table*}
The upper atmospheres of planets can mainly be in two regimes. In the first, the thermosphere is in
hydrostatic equilibrium where the bulk of the upper atmosphere gas below the exobase
can be considered as hydrostatic. In the second, the upper atmosphere may
hydrodynamically expand and the bulk atmospheric particles can escape efficiently as a result of high
stellar XUV fluxes, and/or a weak planetary gravitational field (e.g., Lammer 2012).
Under certain conditions and for certain planetary parameters the upper atmosphere can hydrodynamically
expand, but the bulk velocity may not reach the escape velocity at the exobase level.
In such cases one can expect a hydrodynamically expanding upper atmosphere but the loss
rate results in a less efficient Jeans-type escape, but not in hydrodynamic blow-off
(Tian et al. 2005b; Tian et al. 2008a; Tian et al. 2008b). If the mean thermal energy of
the upper atmosphere gases at the exobase level exceeds their gravitational energy,
blow-off occurs (e.g., \"Opik 1963).

Our simulation domain is limited by a lower boundary
with a temperature $T_0$, density $n_0$ and a thermal velocity
\begin{equation}
v_{\rm 0}=\left(\frac{k T_0}{m}\right)^{0.5}
\end{equation}
at the homopause distance or base of the thermosphere. To estimate the temperature $T_0$ at the
lower boundary one does not need to apply a full radiative transfer calculation because the temperature
in this altitude region differs not greatly from the equilibrium or effective skin temperature
\begin{equation}
T_{\rm eq}=\left[\frac{S(1-A)}{\xi \sigma_{\rm B}}\right]^{0.25},
\end{equation}
of the planet (e.g., Kasting \& Pollack 1983; Bauer \& Lammer 2004; Tian et al. 2005b). The factor
$\xi$ is 4 if a planet rotates fast and 2 for slowly rotating or tidally locked planets.
Because the studied ``super-Earths'' are within Mercury's orbit,
it is likely that they rotate slower compared to the Earth so that $\xi$
may be $<$4. $\sigma_{\rm B}$ is the Stefan-Boltzmann constant, $A$ the albedo and $S$ the
stellar flux density which is the amount of the incoming electromagnetic radiation of the host star
per unit area at the planet's orbit location. For the temperature $T_0$ at the lower boundary of the
studied ``super-Earths'' we use the estimated effective skin temperatures $T_{\rm eq}$ given in Table 1.

For planets close to or inside the
habitable zone of a Sun-type G-star such as Venus or Earth this temperature
is $\sim$230--250 K which is reached at an altitude of about 120 km.

Because of their low mean densities, in the case of
GJ 1214b (Bean et al. 2010; Miller-Ricci \& Fortney 2010; Nettelmann et al. 2010; 2011)
and 55 Cnc e (Demory et al. 2012), it may be possible that these planets have a deep H$_2$O ocean
which is possibly surrounded by a steam atmosphere. Such a scenario is not expected for the Kepler-11
low density ``super-Earths'', these planets are most likely dominated by dense H/He envelopes
(Lissauer et al. 2011; Ikoma \& Hori 2012).
\begin{figure}
\includegraphics[width=0.94\columnwidth]{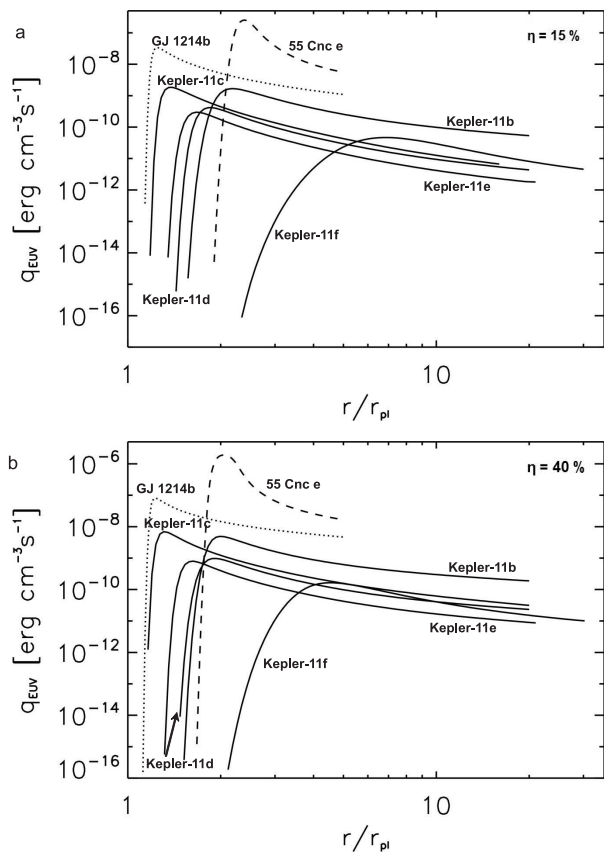}
\caption{Calculated XUV volume heating rates $q_{\rm XUV}$ for 55 Cnc e (dashed line), GJ 1214b (dotted line),
and the 5 Kepler-11 ``super-Earths'' (solid lines) as a function of distance in planetary
radii for $\eta=15$\% (a) and $\eta=40$\% (b), respectively.}
\end{figure}

If a planet has a steam atmosphere the base of the thermosphere could also
act as a kind of cold trap where the saturation H$_2$O mixing ratio reaches
its minimum value. If a planet is dominated by a hydrogen-rich
gas envelope or has different parameters compared to the Earth,
the homopause altitude could also be somewhat higher, but the exobase location or escape rate at the top of
the thermosphere should not greatly differ.

\begin{figure}
\includegraphics[width=0.94\columnwidth]{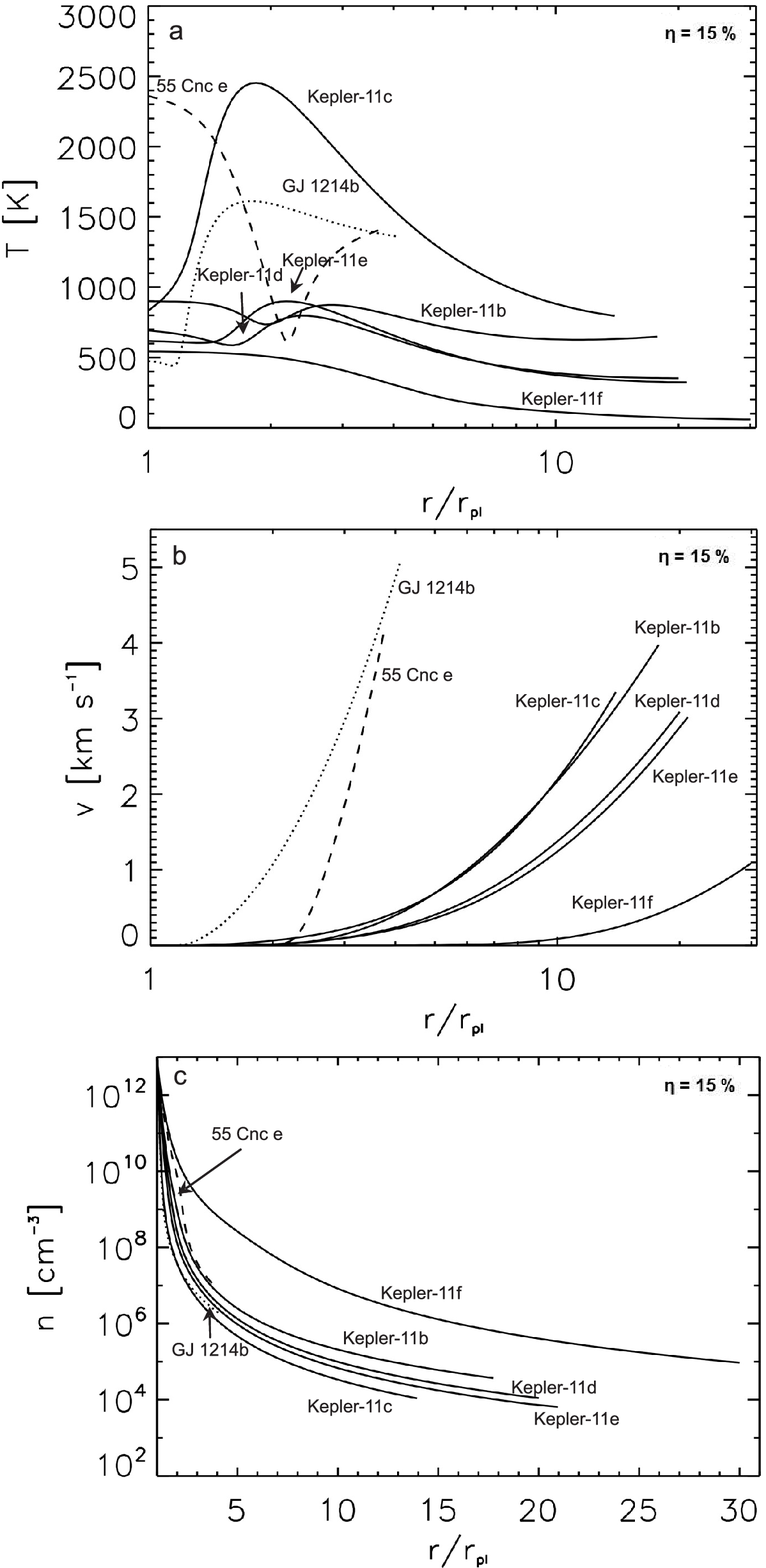}
\caption{Modeled temperature, velocity and number density profiles of 55 Cnc e, GJ 1214b, Kepler-11b-f from the lower thermosphere up to the Roche lobe distance
$r_{\rm L1}$ given in Table 3 for a low heating efficiency with $\eta=15$\%.}
\end{figure}
\begin{figure}
\includegraphics[width=0.94\columnwidth]{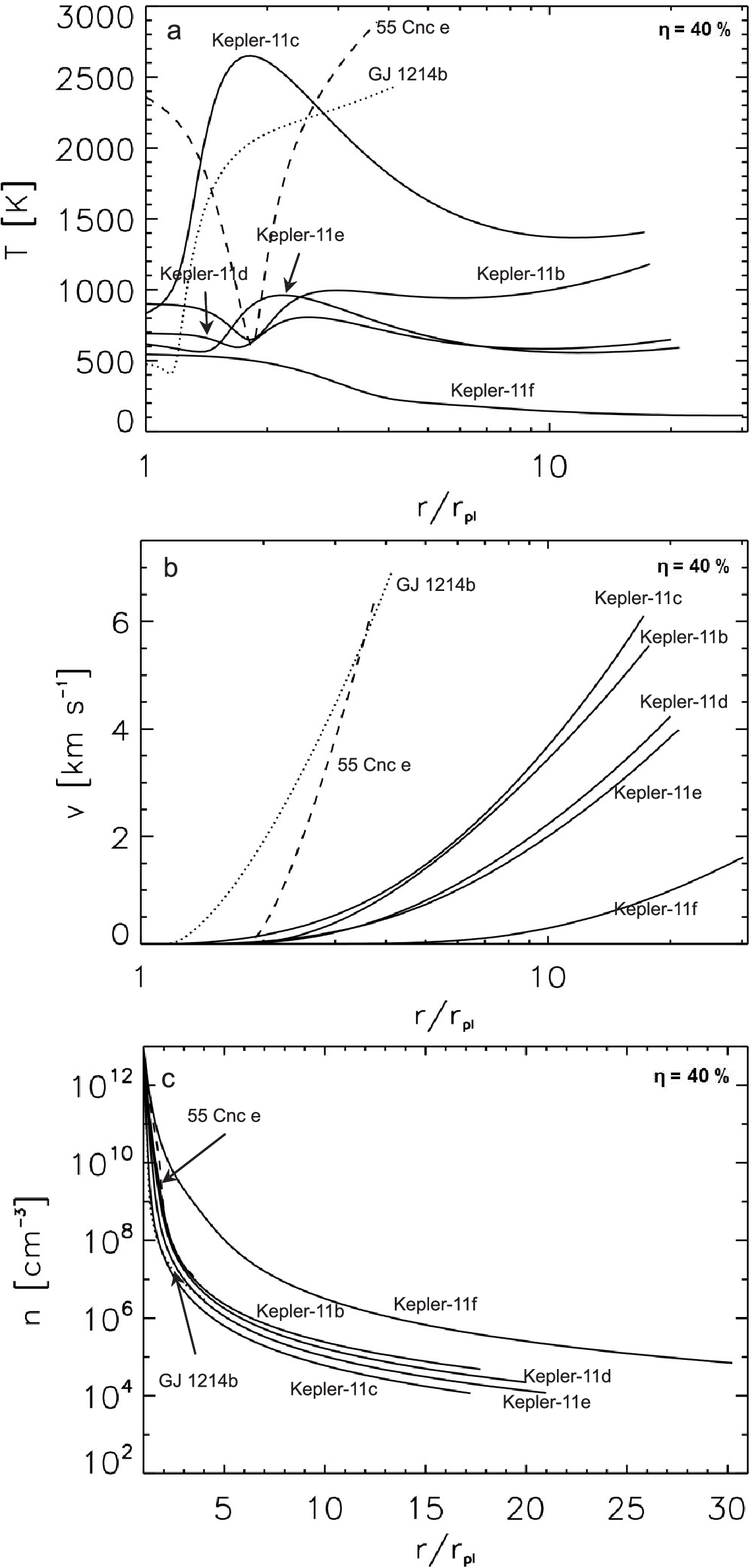}
\caption{Modeled temperature, velocity and number density profiles of 55 Cnc e, GJ 1214b, Kepler-11b-f
from the lower thermosphere up to the Roche lobe distance $r_{\rm L1}$ given in Table 3 for a heating efficiency $\eta=40$\%.}
\end{figure}

\begin{table*}
\begin{center}
\caption{Calculated exobase parameters without Roche lobe effects and thermal escape and mass-loss rates for the five ``super-Earths'' of the
Kepler-11 system, GJ 1214b and 55 Cnc e for heating efficiencies $\eta$ of 15\% (upper part) and 40\% (lower part).}
\begin{tabular}{l|ccccccc}
  Parameter    & Kepler-11b & Kepler-11c & Kepler-11d & Kepler-11e & Kepler-11f & GJ 1214b & 55 Cnc e\\ \hline
Blow-off                &  yes         & no          &  yes          &     yes    &   yes       &    yes      &   yes \\
$r_{\rm exo}$/$R_{\rm pl}$& $\sim 33$   & $\sim 14$  & $\sim 27$  & $\sim 25$  & $\sim 41$ & $\sim 35$ & $\sim 136$ \\
$T_{\rm exo}$ [K]        & $\sim$1170 & $\sim 1000$  & $\sim 415$  & $\sim 430$  & $\sim 50$ & $\sim 4040$ & $\sim 2230$ \\
$n_{\rm exo}$ [cm$^{-3}$]& $\sim 6\times10^{3}$ & $\sim 8 \times 10^3$  & $\sim 5\times10^3$  & $\sim 4\times10^3$  & $\sim 1.6\times10^4$ & $\sim 5\times10^3$ & $\sim 1.75\times10^3$ \\
$L_{\rm th}$ [s$^{-1}$] & $\sim5.6\times10^{31}$ & $\sim1.5\times10^{31}$  & $\sim6.0\times10^{31}$  & $\sim 6.5\times10^{31}$  & $\sim 2.0\times10^{32}$ & $\sim 2.25\times10^{32}$ & $\sim 9.5\times10^{32}$ \\
$\dot{M}_{\rm th}$ [g s$^{-1}$] & $\sim 9.5\times10^{7}$ & $\sim 2.43\times10^{7}$  & $\sim 1.0\times10^{8}$  & $\sim 1.1\times10^{8}$  & $\sim 3.3\times10^{8}$ & $\sim 3.45\times10^{8}$ & $\sim 1.6\times10^{9}$ \\ \hline
Blow-off                &  yes         & yes          &  yes          &     yes    &   yes       &    yes      &   yes \\
$r_{\rm exo}$/$R_{\rm pl}$& $\sim 40$   & $\sim 16$  & $\sim 40$  & $\sim 28$  & $\sim 45$ & $\sim 40$ & $\sim 168$ \\
$T_{\rm exo}$ [K]        & $\sim$1800 & $\sim 1560$  & $\sim 1040$  & $\sim 680$  & $\sim 55$ & $\sim 7125$ & $\sim 5530$ \\
$n_{\rm exo}$ [cm$^{-3}$]& $\sim 6\times10^3$ & $\sim 10^4$  & $\sim 6\times10^3$  & $\sim 5\times10^3$  & $\sim 2\times10^4$ & $\sim 6\times10^3$ & $\sim 1.6\times10^3$\\
$L_{\rm th}$ [s$^{-1}$] & $\sim1.0\times10^{32}$ & $\sim 4.5\times10^{31}$  & $\sim 6.5\times10^{31}$  & $\sim 1.2\times10^{32}$  & $\sim 3.7\times10^{32}$ & $\sim 4.5\times10^{32}$ & $\sim 1.7\times10^{33}$ \\
$\dot{M}_{\rm th}$ [g s$^{-1}$] & $\sim 1.5\times10^{8}$ & $\sim 7.5\times10^{7}$  & $\sim1.0\times10^{8}$  & $\sim2.0\times10^{8}$  & $\sim 6.35\times10^{8}$ & $\sim 7.5\times10^{8}$ & $\sim 2.8\times10^{9}$ \\
\end{tabular}
\end{center}
\end{table*}

In the case of H$_2$/He dominated upper atmospheres there are three processes which can dissociate H$_2$ molecules into hydrogen atoms.
The first process is thermal dissociation which becomes
dominant when the atmospheric temperature reaches values $>$2000 K (Yelle 2004; Koskinen et al. 2010).
The second process is photo-dissociation of hydrogen molecules which becomes dominant when the stellar
EUV flux reaches values which are $>$25 times the present solar value (Koskinen et al. 2010).
The third possibility for the production of atomic hydrogen are photochemical reactions between CH$_4$
and other hydrocarbons taking place below the homopause level (e.g., Atreya et al. 1986). A combination of
photo-dissociation and photochemical reactions is responsible that H atoms are the dominate species in
the upper atmosphere of Jupiter, Saturn, Uranus and Neptune. Because the studied ``super-Earths'' are
much hotter compared to the hydrogen-rich Solar System gas and ice giants, are therefore exposed to higher photon fluxes
and may also contain hydrocarbons in their lower atmospheres (Kuchner 2003) we assume that atomic hydrogen
is the dominant species above their homopause levels.

In case the planet is surrounded by a dense hot steam atmosphere, the stellar XUV radiation dissociates the H$_2$O molecules
above the mesopause via
\begin{equation}
\rm H_2O + h\nu \rightarrow OH + H
\end{equation}
and hydrogen atoms should also be the dominant species in the upper atmosphere (Kasting \& Pollack, Tian et al. 2005; Lammer 2012).
For the  number densities $n_0$ we assume a hydrogen density
of $10^{13}$ cm$^{-3}$ similar as calculated for Saturn's, Uranus' and Neptune's homopause levels (Yamanaka 1995; Atreya 1999). This value is also comparable with the homopause hydrogen density of $5\times 10^{12}$ cm$^{-3}$ which was adopted by Tian et al. (2005b) in a study of a hydrogen-rich early Earth, but slightly lower compared to that of Jupiter.
This density value corresponds to a H$_2$O mixing ratio in Earth's atmosphere of 50 \%
which results, according to Kasting \& Pollack (1983) and Tian et al. (2005b),
also in an upper atmosphere which is dominated by hydrogen.

The upper boundary of our simulation domain is chosen at 45$R_{\rm pl}$, but the results of our hydrodynamic model are considered as accurate only
until the exobase level which is located where the mean free path $l_{\rm c}$ reaches the scale height $H=(kT_{\rm exo})/(m g)$ of the hydrodynamically expanding
upper atmosphere. Above the exobase level which separates the collision dominated thermosphere from the
collision-less exosphere, hydrodynamics are not valid. If $\beta>30$ at the exobase level then an atmosphere is
mainly bound to the planet and one can expect low thermal escape rates. Blow-off happens when $\beta$ becomes $<$1.5
or the upper atmosphere reaches the Roche lobe distance (\"Opik 1963; Chamberlain 1963; Bauer \& Lammer 2004; Erkaev et al. 2007).
For $\beta$ values between $\sim$1.5--30, Jeans escape occurs but the upper atmosphere can expand within the hydrodynamic regime.

Because the classical Jeans formula is based on the isotropic Maxwellian distribution function,
in case if the upper atmosphere expands dynamically but blow-off conditions are not established,
we have the radial velocity of the outward flowing bulk atmosphere at the exobase, and thus a
distribution function which is not isotropic. In such a case we calculate the Jeans-type escape
rate by using a shifted Maxwellian function which is modified by the radial velocity $v$ obtained
from the hydrodynamic code (e.g., Volkov et al. 2011). We show the dynamically expanding upper atmosphere structures only up to the exobase levels.
\section{RESULTS}
The modeled XUV volume heating rates $q_{\rm XUV}$ for the seven ``super-Earths'' with heating efficiencies $\eta=15$\% and 40\% are shown in Fig. 1.
Depending on the planets orbit location, equilibrium or skin temperature, and gravity, the peak of the
XUV flux is deposited close to the planet or at higher altitudes. Corresponding to the volume heating rates
the expected atmosphere temperature, velocity and density profiles as a function of planetary distance without
(Table 2) and with the Roche lobe effects (Table 3) are then modeled.
\begin{table*}
\begin{center}
\caption{Calculated upper atmosphere parameters at the Roche lobe location and thermal escape and mass-loss rates for the studied ``super-Earths''
for heating efficiencies $\eta$ of 15\% (upper part) and 40\% (lower part). Because the exobase level of Kepler-11c does not reach $r_{\rm L1}$, the planet's values at the exobase are
given instead.}
\begin{tabular}{l|ccccccc}
  Parameter              & Kepler-11b & Kepler-11c & Kepler-11d & Kepler-11e & Kepler-11f & GJ 1214b & 55 Cnc e\\ \hline
$r_{\rm L1}/R_{\rm pl}$ & $\sim 17.95$ & $\sim 18.05$& $\sim20.24$ & $\sim 20.96$ & $\sim 30.24$ & $\sim 4.26$ & $\sim 3.82$ \\\hline
$r_{\rm exo}/R_{\rm pl}$ &              & $\sim 14$  &               &            &             &             &               \\
$T_{\rm L1}$ [K]        & $\sim$670 & $T_{\rm exo}\sim 785$  & $\sim 353$  & $\sim 325$  & $\sim 270$ & $\sim 1360$ & $\sim1415$ \\
$n_{\rm L1}$ [cm$^{-3}$]& $\sim 3.7\times10^{4}$ & $n_{\rm exo}\sim 7.0 \times 10^3$  & $\sim 1.0\times10^4$  & $\sim 6.5\times10^3$  & $\sim 9.4\times10^4$ & $\sim 1.8\times10^6$ & $\sim 1.0\times10^7$ \\
$L_{\rm th}$ [s$^{-1}$] & $\sim 7.0\times10^{31}$ & $\sim 2.45\times10^{31}$  & $\sim 6.0\times10^{31}$  & $\sim 6.5\times10^{31}$  & $\sim 2.5\times10^{32}$ & $\sim 4.5\times10^{32}$ & $\sim 9.8\times10^{32}$ \\
$\dot{M}_{\rm th}$ [g s$^{-1}$] & $\sim1.15\times10^{8}$ & $\sim 4.0\times10^{7}$  & $\sim 1.0\times10^{8}$  & $\sim 1.1\times10^{8}$  & $\sim 4.0\times10^{8}$ & $\sim 7.5\times10^{8}$ & $\sim 1.6\times10^{9}$ \\ \hline
$r_{\rm exo}/R_{\rm pl}$ &              & $\sim 16$  &               &            &             &             &               \\
$T_{\rm L1}$ [K]        & $\sim$1190 & $T_{\rm exo}\sim 1415$  & $\sim 650$  & $\sim 590$  & $\sim 110$ & $\sim 2520$ & $\sim 2890$ \\
$n_{\rm L1}$ [cm$^{-3}$]& $\sim 4.8\times10^4$ & $n_{\rm exo}\sim 9.7\times 10^3$  & $\sim 2.2\times10^4$  & $\sim 1.2\times10^4$  & $\sim 7.0\times10^4$ & $\sim 1.5\times10^6$ & $\sim 8.8\times10^6$ \\
$L_{\rm th}$ [s$^{-1}$] & $\sim 1.26\times10^{32}$ & $\sim 7.35\times10^{31}$  & $\sim 1.5\times10^{32}$  & $\sim 1.5\times10^{32}$  & $\sim2.7\times10^{32}$ & $\sim 9.7\times10^{32}$ & $\sim 1.4\times10^{33}$ \\
$\dot{M}_{\rm th}$ [g s$^{-1}$] & $\sim 2.0\times10^{8}$ & $\sim1.3\times10^{8}$  & $\sim 2.5\times10^{8}$  & $\sim 2.5\times10^{8}$  & $\sim 4.5\times10^{8}$ & $\sim 1.5\times10^{9}$ & $\sim 2.3\times10^{9}$ \\
\end{tabular}
\end{center}
\end{table*}
Table 2 shows the calculated exobase parameters $r_{\rm exo}$, $T_{\rm exo}$, $n_{\rm exo}$, the thermal escape rates
$L_{\rm th}$ and mass-loss rates $\dot{M}_{\rm th}$ for the five ``super-Earths'' in the Kepler-11 system, GJ 1214b and
55 Cnc e for heating efficiencies of 15\% and 40 \%, respectively. For example, for Kepler-11f the exobase would be located at $\sim 117 R_{\rm \oplus} \sim 10R_{\rm Jup} \sim 1 R_\odot$. In the case of GJ 1214b which orbits around a M4.5 dwarf star with a radius of $\sim 0.21R_\odot \sim 23R_{\rm \oplus}$ the exobase would be located at $\sim 4.0R_{\rm GJ 1214}$. This comparison indicates that the inflated upper atmosphere of GJ 1214b would cover the whole star during a transit. In the case of 55 Cnc e the expanding bulk atmosphere
would reach escape velocity and hence blow-off conditions far below the estimated exobase level, so that the exobase location
in that particular case is not relevant for the estimation of the thermal escape rate. For a heating efficiency $\eta$ of 15\% blow-off would occur at
$\sim20R_{\rm pl}$ and for $\eta=$40\% at $\sim14R_{\rm pl}$. Due to the lower gravity the combination of XUV heating
and thermal expansion the exobase levels of Kepler-11b, Kepler-11d, Kepler-11e, Kepler-11f, GJ 1214b and 55 Cnc e move to distances which are far
beyond the $L_1$ point $r_{\rm L1}$. Therefore, one can not neglect Roche lobe effects for these particular ``super-Earths''.
The thermal atmospheric escape rates for most planets are $\sim 2$ times higher if the heating efficiency $\eta$ is 40 \%.
For $\eta=40\%$ all ``super-Earths''
are in the blow-off regime. For the lower heating efficiency of 15\%, only Kepler-11c is not in blow-off but experiences strong Jeans-type
escape at the exobase level from its non-hydrostatic expanded upper atmosphere which is about a factor of 0.65 lower compared to the
hydrodynamic outflow rate at the exobase level. The usual Jeans approach which assumes zero bulk velocity would yield a value which is a factor of 0.145 lower.

Another interesting effect on the exobase temperature can be seen at
the most distant ``super-Earth'', Kepler-11f where $T_{\rm exo}$ cools adiabatically to a very low
value of $\sim 55$ K.
\begin{table*}
\renewcommand{\baselinestretch}{1}
\caption{Estimated hydrogen loss rates from Eq. (12) for energy-limited escape ($\eta=100$\%) and heating
efficiencies of 40\% and 15\%.}
\begin{center}
\begin{tabular}{l|ccc}
Exoplanet     &  $\eta=100$\%: $L_{\rm th}$ [s$^{-1}$] & $\eta=40$\%: $L_{\rm th}$ [s$^{-1}$]  & $\eta=15$\%: $L_{\rm th}$ [s$^{-1}$] \\\hline
Kepler-11b &  $\sim 6.0\times 10^{32}$ &   $\sim 2.4\times 10^{32}$   &   $\sim 9.0\times 10^{31}$   \\
Kepler-11c &  $\sim 5.9\times 10^{32}$ &   $\sim 2.3\times 10^{32}$   &    $\sim 8.9\times 10^{31}$  \\
Kepler-11d &  $\sim 7.5\times 10^{32}$ &   $\sim 3.0\times 10^{32}$   &    $\sim 1.1\times 10^{32}$  \\
Kepler-11e &  $\sim 8.0\times 10^{32}$ &   $\sim 3.2\times 10^{32}$  &   $\sim 1.2\times 10^{32}$   \\
Kepler-11f &  $\sim 3.5\times 10^{32}$ &   $\sim 1.4\times 10^{32}$  &   $\sim 5.3\times 10^{31}$   \\
GJ 1214b   &  $\sim 4.6\times 10^{33}$ & $\sim 1.9\times 10^{33}$  &  $\sim 7.0\times 10^{32}$  \\
55 Cnc e   &  $\sim 1.7\times 10^{34}$ &   $\sim 7.0\times 10^{33}$  &   $\sim 2.6\times 10^{33}$   \\
\end{tabular}
\end{center}
\normalsize
\end{table*}
However, as mentioned above it is important to note that the results in Table 2 neglect the Roche lobe and
its escape-enhancing effects. If we consider this more realistic scenario and compare the exobase levels of the studied ``super-Earths''
with the location of the $L_1$ point $r_{\rm L1}$ shown in Table 3 one can see that all ``super-Earths'' except Kepler-11c experience escape due to Roche lobe overflow.
This geometrical blow-off is most extreme for 55 Cnc e and is also effective for GJ 1214b for which the distances
of the respective L1 points $r_{\rm L1}$ are about $4R_{\rm pl}$.
Due to the low gravities, the enormous expansions of the upper atmospheres and the relatively close distances to their host stars, one cannot neglect the Roche
lobe effects. Table 3 shows the calculated parameters, but with Roche lobe effects included at the distance of the corresponding L1 points. In such a case the atoms escape
as they reach and overflow the Roche lobe.

If we compare the escape rates which neglect the Roche lobe effects shown in Table 2 with those of Table 3 where Roche lobe effects are included, one can
see that the stellar tidal forces enhance the atmospheric escape. The escape rates are influenced more strongly at the close-in ``super-Earths''
GJ 1214b and 55 Cnc e compared to the Kepler-11 planets where $r_{\rm L1}$ is further away from $r_{\rm pl}$.
One can also see that the temperatures are different at the exobase level for Kepler-11c
without (Table 2) and with Roche lobe effects (Table 3). The reason is that with the Roche lobe effects, the radial flow velocity
is larger, which results in a larger adiabatic cooling. One should also note that a number density which is less or larger than the
assumed $10^{13}$ cm$^{-3}$ at the homopause level would result in a lower or higher escape rate.

Figs. 2 and 3 have the Roche lobe effects included and show the calculated temperature, velocity and density profiles as a function of planetary radii
from the homopause distance up to the Roche lobe and in the case of Kepler-11c up to the exobase level.
In Fig. 2 we consider a low heating efficiency $\eta$ of 15\%, i.e. the availability of IR-cooling molecules in the
upper atmosphere and in Fig. 3 we use a heating efficiency $\eta=40$\%
as expected for hydrogen-rich hot gas giants (e.g., Yelle 2004; Koskinen et al. 2012).

One can see in Figs. 2a and 3a that the temperature structure of these ``super-Earth'' upper atmospheres is determined by a
complex interplay between the XUV flux, the planet's mean density, and its skin or equilibrium temperature
$T_0\approx T_{\rm eq}$. In the case of the hottest ``super-Earth'' 55 Cnc e the maximum XUV flux
is deposited at higher altitudes (see also Fig. 1), therefore, one can see that the temperature first decreases until the
XUV radiation heats the atmosphere and expansion related to adiabatic cooling becomes relevant. For Kepler-11c
which is the heaviest ``super-Earth'' with a cooler skin temperature the XUV flux is deposited at a closer
distance so that the atmosphere is heated to a temperature peak which is above 2500 K. Then adiabatic cooling
due to hydrodynamic expansion cools the upper atmosphere at large distances to temperatures below 1500 K
for an $\eta$ of 40\% and $<$1000 K if $\eta=15$\%. One can also see that for the ``super-Earth''
which is furthest away from its host star, namely Kepler-11f, the temperature decreases until the upper atmosphere
reaches the Roche lobe distance. The XUV flux is too low to raise the temperature around the altitude where the
maximum energy is deposited. Due to this different behavior, Kepler-11f is the only ``super-Earth'' in our sample
where the energy-limited formula in Eq. (12) underestimates the escape rate.
From Figs. 2b and 3b one can see that the expanding hydrogen atoms have velocities
of a few km s$^{-1}$ but $<$10 km s$^{-1}$. Figs. 2c and 3c show the corresponding number density profiles.

After the investigation of the blow-off regimes and upper atmosphere structures of the seven ``super-Earths''
we compare the calculated thermal escape rates with escape estimates from the simple energy-limited escape formula
(e.g. Watson, Donahue, \& Walker 1981; Hunten, Pepin, \& Walker 1987), but introduce a heating efficiency $\eta$ and
a potential energy reduction factor due to do the stellar tidal forces (Erkaev et al. 2007)
\begin{equation}
K=1-\frac{3}{2(r_{\rm L1}/R_{\rm pl})} + \frac{1}{2(r_{\rm L1}/R_{\rm pl})^3}.
\end{equation}
The equation can then be written as
\begin{equation}
L_{\rm th}\approx\frac{3\eta I_{\rm XUV} F_{\rm XUV}}{4 G \rho_{\rm pl} m_{\rm H}K},
\end{equation}
where $I_{\rm XUV}$ is the XUV enhancement factor given in Table 1,
$F_{\rm XUV}$ is the present time solar XUV flux at 1 AU, $G$ is
Newton's gravitational constant, $\rho_{\rm pl}$ is the planetary density and $m_{\rm H}$ is the mass of a hydrogen atom.
The results are shown in Table 4 for the energy-limited case with $\eta=100$\% and heating efficiencies of 40\% and 15\%.

The Roche lobe potential energy reduction factor $K$ plays only a role for GJ 1214b and 55 Cnc e because the $L_1$ point of the Roche lobe is much closer
to the planetary surfaces as compared to the other ``super-Earths''. $K$ is 1.53 for GJ 1214b and 1.63 for 55 Cnc e and
$\sim 1$ for all five Kepler-11 planets. If we compare the thermal escape rates estimated by the simple Eq. (12) with
those obtained by the hydrodynamic model in Table 3, one can see that, besides of Kepler-11f, the atmospheric escape rates are only slightly overestimated.
A difference of a factor 2--3 is certainly
within all the other uncertainties such as the XUV flux enhancement. For that reason,
our study indicates that the simple formula given in Eq. (12) is useful for mass-loss estimates of hydrogen-rich
exoplanets during evolutionary time scales, if the particular planet is exposed to a high XUV
flux and/or is located at close orbital distance.
However, Eq. (12) should not be used if the atmosphere of a planet is not
within the blow-off regime.
\begin{figure}
\includegraphics[width=0.94\columnwidth]{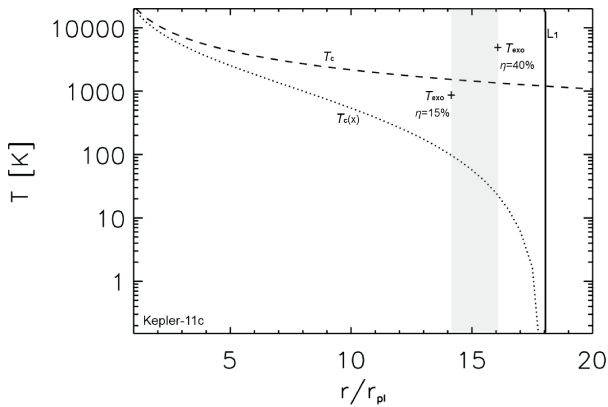}
\caption{Comparison of the critical temperature $T_{\rm c}$ without (dashed line) and $T_{\rm c}(x)$ with Roche lobe effects (dotted line)
at Kepler-11c as a function of planetary radii. The symbol ``+'' corresponds to $T_{\rm exo}$ related to the low and higher $\eta$ values.
The vertical solid line corresponds to the Roche lobe distance for the $L_{1}$ point and
the shaded area marks the exobase levels of Kepler-11c for heating efficiencies $\eta$ of 15\% and 40\%, respectively.}
\end{figure}
By comparing the exobase location with the distance of the $L_1$ point of the most massive ``super-Earth'' Kepler-11c, one can
see that the exobase does not reach the Roche lobe distance. By considering the Roche lobe effects one can see from Table 3 that
the exobase temperature $T_{\rm exo}$ of Kepler-11c is $\sim$785 K for a heating efficiency $\eta$ of 15\% and $\sim$1415 K if
$\eta$ is 40\%. The classical critical temperature $T_{\rm c}$ for blow-off  can be written as (e.g. Hunten 1973)
\begin{equation}
T_{\rm c}=\frac{2 G m_{\rm H} M_{\rm pl}}{3 k r_{\rm exo}}.
\end{equation}
By using the corresponding parameters for Kepler-11c given in Tables 1 and 3, $T_{\rm c}$ is $\sim$1490 K for $\eta$=15\% and $\sim$1350 K
if $\eta$=40\%. The corresponding exobase temperatures $T_{\rm exo}$ for the particular planet are $\sim$785 K and 1415 K, respectively. By using this
criteria one finds that the upper atmosphere of Kepler-11c would be in the blow-off stage for the higher heating efficiency but not for
the lower value of 15\%. However, Erkaev et al. (2007) showed that stellar tidal forces influence also the
critical temperature $T_{\rm c}$ so that under certain conditions blow-off can occur.
If one follows the derivations related to the potential energy difference
between the exobase level and the Roche lobe, given in Erkaev et al. (2007) we
obtain the following equation
\begin{equation}
T_c(x)=T_{\rm Jup}\left(\frac{M_{\rm pl} R_{\rm Jup}}{M_{\rm Jup} R_{\rm pl}}\right)\frac{K(x_{L{_1}}/x)}{x},
\end{equation}
with
\begin{equation}
x=\frac{r_{\rm exo}}{R_{\rm pl}}\hspace*{0.3cm}{\rm and}\hspace*{0.3cm}x_{L_{1}}=\frac{r_{L1}}{R_{\rm pl}}.
\end{equation}
Here $T_{\rm Jup}\sim 1.45 \times 10^5$ K is the critical temperature for the
onset of blow-off at the radius of Jupiter (i.e. when $r_{\rm exo}=R_{\rm pl}$),
$M_{\rm Jup}$ and $R_{\rm Jup}$ are the mass and radius of Jupiter. Fig. 4 compares
the temperature profiles $T_{\rm c}$ and $T_{\rm c}(x)$ of Kepler-11c as a function of
planetary radius without and with Roche lobe effects.
From this one can see that Roche lobe effects reduce the critical temperature $T_{\rm c}(x)$
at the exobase level to $\sim$80 K for a heating efficiency $\eta$ of 15\%, resulting
in an environment where Kepler-11c's upper atmosphere will experience blow-off.
But is interesting to note that Eq. (12) overestimates in that particular case the
atmospheric escape rate by a factor of $\sim$3.6.

Compared to the mass-loss estimations of ``hot Jupiters'' such as HD 209458b which are in the order of $\sim$2--5$\times 10^{10}$ g s$^{-1}$
(e.g. Yelle 2004; Tian et al. 2005a; Garc\'{\i}a Mu\~{n}oz 2007; Penz et al. 2008; Lammer et al. 2009a; Koskinen et al. 2012) the mass-loss of the studied
``super-Earths'' is $\sim$10--100 times lower. The main reasons are that the studied ``super-Earths'' are exposed to lower XUV fluxes and
the mean density of HD 209458b is lower.
\section{CONCLUSION}
We studied the blow-off regimes of seven hydrogen-rich ``super-Earths'' with known sizes and masses by applying a
1D hydrodynamic upper atmosphere model which includes tidal forces. The upper atmosphere temperature profiles of these
hydrogen-rich ``super-Earths'' are determined by a complex interplay between the equilibrium or skin temperature of the particular
planet, the homopause number density, the stellar XUV flux, the height where the maximum XUV energy is deposited, adiabatic cooling
the mean density of the particular planet, and the Roche lobe distance.
The upper atmospheres of Kepler-11b, Kepler-11d, Kepler-11e, Kepler-11f, GJ 1214b and 55 Cnc e expand up to the $L_1$ points
of their Roche lobes, which enhances their atmospheric escape rates. Kepler-11c is the only ``super-Earth'' in our sample where
its exobase level does not reach the Roche lobe. The thermal mass-loss rates of GJ 1214b and 55 Cnc e are $\sim$10 times lower
and about 100 times lower for the
Kepler-11 ``super-Earths'', respectively, compared to a typical ``hot Jupiter'' such as HD 209458b. By comparing the
escape rates obtained by the hydrodynamic upper atmosphere model with the modified energy-limited equation given in Erkaev et al. (2007)
and Lammer et al. (2009) we found that besides of Kepler-11c and Kepler-11f and 55 Cnc e
the escape rates are overestimated by about a factor two, which lies within all other parametrical uncertainties.
However, our study also indicate that the energy-limited equation should only be applied for hydrogen-rich exoplanets
which have a hot skin or equilibrium temperature and/or are exposed to high XUV flux values.
From our results one can expect that the mass of the investigated ``super-Earths'' will not be affected
strongly by atmospheric mass-loss during their remaining lifetimes.

\section*{ACKNOWLEDGMENTS}
N. V. Erkaev, K. G. Kislyakova, M. L. Khodachenko, \& H. Lammer acknowledge the support by the FWF NFN project S116 ``Pathways to Habitability:
From Disks to Active Stars, Planets and Life'', and the related FWF NFN subprojects, S116 606-N16 ``Magnetospheric Electrodynamics
of Exoplanets'' and S116607-N16 ``Particle/Radiative Interactions with Upper Atmospheres of Planetary Bodies Under Extreme Stellar Conditions''.
K. G: Kislyakova, H. Lammer, \& P. Odert thank also the Helmholtz Alliance project ``Planetary Evolution and Life''.
M. Leitzinger and P. Odert acknowledge support from the FWF project P22950-N16. N. V. Erkaev acknowledges support by the
RFBR grant No 12-05-00152-a. The authors also acknowledge support from the EU FP7 project
IMPEx (No.262863) and the EUROPLANET-RI projects, JRA3/EMDAF and the Na2 science working group WG5. The authors thank the
International Space Science Institute (ISSI) in Bern, and the ISSI team ``Characterizing stellar- and exoplanetary environments''.
Finally, we thank an anonymous referee for interesting suggestions and recommendations which helped to improve the article.

\end{document}